# Water Need Models and Irrigation Decision Systems

**Meriç Çetin*, Senem Yıldız**, Selami Beyhan***,**

*Pamukkale University, Department of Computer Engineering, Kinikli Campus, 20070, Denizli, Turkey (mcetin@pau.edu.tr).

**Izmir Democracy University, Department of Electrical and Electronics Engineering, Uckuyular Dist., Gursel Aksel Blv., 35140 Karabaglar, Izmir, Turkey(senemyildiz607@gmail.com).

***Izmir Democracy University, Department of Electrical and Electronics Engineering, Uckuyular Dist., Gursel Aksel Blv., 35140 Karabaglar, Izmir, Turkey (selami.beyhan@idu.edu.tr).

## Abstract

Irrigation decision systems and water need models have been important research topics in agriculture since 90s. They improve the efficiency of crop yields, provide an appropriate use of water on the earth and so, prevent the water scarcity in some regions. In this paper, a comprehensive survey on water need models depending on crop growth and irrigation decision systems has been conducted based on mathematical maodelling. The following outcomes and solutions are the main contributions. Crop growth models and correspondingly water need models are suffer from un-modeled dynamics of the environment and lack of sensory devices. Literature review with the latest developments on water need models, irrigation decision systems, applied control methods and discussions are expected to be useful for the future strategies.

## 1 Introduction

Considering the climate change and agricultural practices, applications related to water conservation have become remarkable. Nowadays, the decrease of freshwater water resources due to global population growth, climate factor, industry, improper irrigation and agriculture policies have contributed to the development of precise and intelligent automatic irrigation technologies. However, it is thought that the mathematical models that will determine the water need are not precise enough and the developments in the irrigation systems are based on uncertainties due to many disturbing factors such as inability to make exact precipitation prediction, quality of seed, minerals in the water, weather forecast and momentary changes in the weather. In order to improve irrigation control, the scientific community and farmers do the methods developed in their joint studies in controlled agricultural lands. Improvements have started to give more results with the reduction of external factors and the use of more measurements in controlled land. Over the last 50 years, various methods have been proposed for the rational usage and sustainability of resources in global agriculture [1]. Effective irrigation planning is very important in terms of sustainability of global food system, maintain of environment quality, production using less resources in the agriculture industry [2]. Therefore, compared to other systems, it is thought that water will be properly distributed and used with irrigation systems that are automated depending on time, soil moisture content or temperature.

# 2 Modeling the Water Need of Crops and Irrigation Systems

Deriving a difference equation or differential equation that represents the system dynamics is called modeling [3], where accurate control or estimation tasks are performed based on the constructed system model. Modeling is performed within the framework of physical laws (mathematical modeling) or by emprical methods. System identification is the empirical modeling task obtained as a white, gray or black box by a limited experimental data from a particular dynamic system [4]. System identification is usually consulted when the all inputs and disturbances of the system is unknown. The success of system identification depends on many variables. The most important of these are the designed model, the data, the optimization of the model parameters and the correct determination of all possible inputs. In irrigation systems, mathematical modeling was generally used and some models were developed [5]. It is difficult to test these models directly as there are too many inputs that affect the irrigation mechanism in plant growing. Instead of that designed models have been evaluated within the control methods according to the correct irrigation and the success achieved [6, 7]. In later studies, the system model was constructed using structures such as Artificial Neural Networks (ANNs)-based on identification. In addition, it is thought that Machine Learning (ML) models, which have become more popular in recent years, will increase the identification success and control performance.

The water resources are decreasing today due to population growth, climate factor, industry and improper agricultural applications. The concept of water balance is related to the calculation of all the amount of water added, extracted and stored for a crop in a given field and in a given time period. Proper application of water is called irrigation in order to meet the water requirement of the plant or crop on time. It should be taken into consideration that each crop needs different planning for irrigation and factors such as rainfall and groundwater affect the water demand. The precise and smart automatic irrigation technologies developed in this context are based on various mathematical models that provide water balance by using information such as evaporation, percolation and field surface flow in feedback. Various parameters (like geo-hydrological, ecological, economic, political etc.) should be integrated into these models in order to find proper solutions to water resources problems [8]. Many of the proposed models are designed for use in predictive [9-11], descriptive integrated [12-14] or combined with machine learning methods [15-17]. In this section, irrigation system models that are widely used in the literature are mentioned.

In [18], evapotranspiration analysis is focused such that the fundamental definitions about evaporation, transition, reference crop evapotranspiration $(ET_0)$, crop evapotranspiration under standard conditions $(ET_c)$ are mentioned and explained in detail. The evapotranspiration consists of two types as lost of water in soil and vapour in crop fields. Weather, crop characteristics, administrative and environmental factors are parameters that affect evapotranspiration. Factors such as radiation, crop type, air temperature, humidity, wind speed, soil salinity, soil fertility, fertilizer application can be listed as examples of these parameters. In [18], main evapotranspiration expression is constructed as Penman Monteith equation

$$\lambda ET = \frac{\zeta(r_n-\alpha)+\sigma_a h_p \frac{(v_s-v_a)}{r_a}}{\zeta+\mu(1+\frac{r_s}{r_a})} \tag{1}$$

where $r_n$ is the net radiation, $\alpha$ is the soil heat flux, $(v_s - v_a)$ is the vapour pressure deficit of the air, $\sigma_a$ is the mean air density at constant pressure, $h_p$ is the specific heat of the air, $\zeta$ is the slope of the saturation vapour pressure temperature relationship, $\mu$ is the psychrometric constant, $r_s$ is surface resistance and $r_a$ is the aerodynamic resistance, respectively. Regarding to this equation, the irrigation water requirement is able to be defined as the difference between the crop water requirement and effective precipitation. In [19], combining the Wireless Sensor Network (WSN) and automation technology, improvement of economic productivity has been aimed in viticulture. Modeling was mentioned and explained as relationship between soil moisture and applied water. NICTOR WSN platform and some kinds of sensors for collecting the data in 20cm depth such as Theta probe, Echo probe have been used. The equation is

$$\dot{\theta}(t) = c'_1 I(t) - c'_2 T_e(t) - c'_3 \theta(t), \tag{2}$$

where $I$ is the inflow term, $T_e$ is the box temperature, $\theta$ is the soil moisture and, $c_i{}'$s are unknown parameters which will be estimated. By sampling the equation in interval $T$

$$\theta(k+1) = c_1 I(k-\tau) - c_2 T_e(k) - c_3 \theta(k), \tag{3}$$

where $k$ indicates the sample time and $\tau$ is the start time. The results say that a simple irrigation model can describe the dynamic behavior of the relationship of soil moisture and water usage. Furthermore, this model enables the significant amount of reduction of water usage and economic benefits. [20] mentions the theoretical analysis of the water balance method, which includes additional water and crop transpiration measurement. To save water, an irrigation control system constructed on $ET$ (crop transpiration) has been designed using the Internet of Things (IoT). Irrigation system has been characterized by the water balance method as follows

$$m = W - P - G + ET_c, \tag{4}$$

where $m$ is the irrigation volume, $W$ is the field capacity, $P$ is the effective rainfall, $G$ is the groundwater recharge and $ET_c$ is the crop transpiration, respectively. $ET_c$ is constructed by the product of $ET_0$ and crop transpiration coefficient $K_c$. The hardware components in the field and the irrigation server constitute the intelligent parts of the system. Hardware consists of a controller and weather station. Furthermore, controller communicates with server through smart phones. The system software uses data processing, control and decision-making units and a control website to monitor the data and model. In [20], problems such as difference in soil moisture and delay in irrigation decisions have been considered.

An irrigation water management model was presented for water supply taking into account rainfall and irrigation water in [21]. In this model, the sum of rainfall and irrigation is evaluated as input signal. According to the model, the output signals are crop evapotranspiration, surface flow, groundwater and leakages. The water balance model in [21] is given by Eq. (5).

$$\begin{aligned} S_i \quad &= S_{i-1} + Pr_i + Ch_i - ET_i - DR_i - DF_i - Rh_i, \\ &\text{IF} S_{i-1} + Pr_i < Ni, \text{THEN} IR_i > 0 \\ &\text{IF} S_{i-1} + Pr_i \geq Ni, \text{THEN} IR_i = 0 \\ &N_i = ET_i + DF_i + Rh_i \end{aligned} \tag{5}$$

where the terms $S_i, Pr, Ch, ET, DR, Rh, N, IR$ and $R_i$ represent field storage, rainfall, channel irrigation water applied, crop evapotranspiration, surface runoff-overflow from field, vertical percolation, lateral seepage inflow, field losses from the system, irrigation water requirement and channel water volume, respectively. The time period is defined with suffix $i$. According to the model in Eq. (5), the decision to irrigate or not is determined as: $\text{IF} Ri_i \geq IR_i, \text{THEN} Ch_i = IR_i$, $\text{IF} Ri_i < IR_i, \text{THEN} Ch_i = Ri_i$ and $IR_i = St_i - (S_{i-1} + P_i) + N_i$ . With this proposed smart irrigation management, field irrigation water depth was automatically identified and field water gates were controlled through IoT.

In [22, 23], a new point of view is developed by adding more parameters to the irrigation model SPAW-IRRIG (Soil-plat-air-water). These parameters are economical aspect, knowledge about yield responding to water, irrigation scheduling. The main aim is to design a different methodology to increase the efficiency of seasonal irrigation system by using economical aspect. This model is the crop simulation model which determines the climatic conditions, soil properties, agronomic characteristics and production decisions. Estimating the water usage of crops and combination of yield's impacts are the managing points by using three step procedure in calculations. [24] introduces a model for water usage in growing-season in five different landing area. For a chosen land area, the program iterates the almost all possible combination of water usage. Inputs of these system are pricing, production cost, irrigation cost and maximum yields. Influences of inputs are observed on others when the net return is able to be evaluated after many iterations of the model. Crop water allocation model evaluates the net return for every combinations of water usage in chosen land area. The equation of net return is

$$Netreturn = (comodityprice) * (yield) - (irrigationcost + productioncost), \quad (6)$$

where commodity prices are user inputs. The aim is calculate and reach the optimum allocations in limited water situations when the net return is maximized. In [25], basin irrigation system has been mentioned and developed. The basin irrigation is clearly explained as closed basins and sequential basins by modeling the irrigation controller. The modeling consists of two parts as hydrodynamic basin irrigation models which means that overland flow explained by hydrodynamic flow equations which are

$$\frac{dh}{dt} + \frac{d(hu)}{dx} + \frac{d(hv)}{dy} + 1 = 0 \quad (7)$$

$$\frac{d(uh)}{dt} + \frac{d(u^2h)}{dx} + \frac{d(uvh)}{dy} + gh\frac{dH}{dx} + ghS_f x = 0 \quad (8)$$

$$\frac{d(vh)}{dt} + \frac{d(uvh)}{dx} + \frac{d(uvh)}{dy} + gh\frac{dH}{dy} + ghS_f y = 0 \quad (9)$$

The model name comes from the two dimensional zero-inertia approximation. In addition to this model, same numerical solutions which includes governing equations, spatial discretisation of basin layout domain and in corporation of inflow and outflow boundaries are handled in detail.

A simplified water balance model is used for irrigation control in [26]. When $D$ is given as root zone soil moisture deficit at the current time step, its value in the next time step is given as follows.

$$D^+ = D + E^* - P^e - I^e \quad (10)$$

where $E^*$, $I^e$, $P^e$ are crop evapotranspiration, irrigation amount and effective rainfall amount, respectively.

The decision structure in [27] has been proposed to perform irrigation planning. Based on soil water balance and crop yield, this model uses climate, crop and soil data as input. The soil water balance expression is written as

$$SMD_{i,j} = SMD_{i-1,j} + ET_{i,j} + DP_{i,j} - I_{i,j} - R_{i,j} \quad (11)$$

where $SMD$ is total soil moisture depletion in the root zone, $ET$ is the crop evapotranspiration, $DP$ is the deep percolation, $I$ is the net irrigation amount, $R$ is the effective rainfall, $i$:time-index and $j$:space index (fields), respectively [27]. The crop stress factor ($K_s$) determines the effect of water stress on crop evapotranspiration. The crop stress factor is estimated using a linear or a logarithmic function. In addition, crop yield can be estimated from seasonal crop evapotranspiration values as

$$\frac{Y_a}{Y_m}(1 - b_e) + b_e(\frac{ET_s}{ET_m}) \quad (12)$$

where $Y_a$ is the actual crop yield, $Y_m$ is the maximum attainable crop yield. $b_e$ is the difference between crop evapotranspiration and yield coefficient. $ET_s$ is the seasonal evapotranspiration and $ET_m$ is the maximum seasonal evapotranspiration. A comprehensive seasonal furrow irrigation system that estimates soil moisture levels based on water balance has been proposed in [28]. The model also includes the soil moisture model, irrigation hydraulic model, crop yield model and economic optimization model. Advance times and volume balance are predicted with the hydraulic irrigation model. Besides, the following expression is given as a crop yield estimation tool

$$Y_a = Y_{max}\prod_{j=1}^{4}\left[1 - k_{Y_j}\left(1 - \frac{ET_j}{ET_{M_j}}\right)\right] \quad (13)$$

where $j$:space index, $Y_{max}$ is the maximum yield in the past and $k_{Y_j}$ is the coefficient of crop response. The relationship between accumulated real and maximum evapotranspiration is $\frac{ET_j}{ET_{M_j}}$ [28]. After the irrigation season is over, the net profit maximization is calculated by subtracting the product of expected crop yield ($Y$) and product sales price ($P_C$) from the irrigation cost function ($f_C$). The model used in [29] calculates crop yield as in Eq.(13) by taking into account the deficiency coefficient ($C_{dmi}$) in the root as $\frac{Y_a}{Y_{max}} = \prod_{i=1}^{n}(1 - (k_{Y_i}(C_{dmi}(1-p)_i)))$. $(1-p)_i$ is the crop evapotranspiration fraction for each of the growth. In [29], cost is modeled by crop yield for the gross margin calculation.

$$G_m = Y_r P_{pp} + Y_{sp} P_{sp} + C_p - P_c - C_i D_{gd} \quad (14)$$

where $G_m$ is the gross margin, $Y_r$ is the real harvested yield, $P_{pp}$ is the selling price of product, $Y_{sp}$ is the commercial by crop yield, $P_{sp}$ is the selling price of the by-product, $C_p$ is the payments made from compensatory subsidy, $P_c$ is the global production costs for actual yield, $C_i$ is the cost of irrigation water application and $D_{gd}$ is the gross depth, respectively.

In [30], a pond-irrigation model has been proposed to predict crop irrigation demand and pond water amount. Surface water flow into a pool is defined as $R = \frac{(P-0.2S)^2}{(P+0.8S)}$. $R$ is the surface runoff rate, $P$ is the rainfall rate and $S$ is the watershed retention parameter. The pond water evaporation is $E_p = K_p\frac{R_r}{\zeta}$ where $E_p$ is the evaporation from pond water, $K_p$ is the coefficient, $R_r$ is the solar radiation and $\zeta$ is the latent heat of vaporization. According to Darcy's Law, the lateral infiltration of water from soil into the pool or in the opposite direction is defined as: $D_l = A_l K_h \frac{h_{soil} - h_{pond}}{l}$. $D_l$ is the lateral infiltration rate, $A_l$ is the flow area perpendicular to $l$, $K_h$ is the saturated hydraulic conductivity, $h$ is the hydraulic head and $l$ is the flow path length, respectively. Table 0 summarizes the models used for irrigation systems in the literature.

## 2.1 An Assessment for Modeling Water Needs of Irrigation Systems

It is known that most of the global water consumption is used in irrigation. Therefore, modeling of irrigation water needs has become necessary in order to evaluate the future water situation [31]. Today, the problem of water scarcity due to climate change has also arisen. Most of the irrigation system models used in the literature do not consider this climate change factor or meteorological conditions effect as a parameter or input. Conversely, future water needs need to be assessed with more precise models that take into account the climatic factor and weather forecast. Improvements in climate models that exhibit both regular and chaotic behavior lead to positive results in short-term weather forecasts [32]. Despite these advances, there are still many uncertain parameters in long-term predictions. It is expected that there will be a significant performance increase in irrigation systems as a result of reducing existing uncertainties with identification methods and increasing the accuracy of irrigation models. There are applications where time series are needed to analyze the effect of climate variables on irrigation requirements. For these applications, factors such as precipitation, temperature, radiation, sunshine, snow-melt and the number of wet or dry days are effective [33]. By using such an irrigation model, water scarcity can be evaluated according to current values and demand. The water level in the soil is checked by taking into account the meteorological conditions of the next day. Thus, it can be determined dynamically whether the irrigation system will work or not. In addition, irrigation water requirements to ensure optimum crop growth can be calculated in such a way that evapotranspiration occurs at the potential rate.

## 2.2 Recommendations for Water Management Strategies in Irrigation Decision Systems

The development in irrigation systems, which has an important role in agricultural development, should be evaluated together with the increase of the population. The sustainable use of water in agriculture has priority in terms of conservation of resources, environmental awareness, technology adaptation and economics [2]. There is no fixed criterion for irrigation according to differences in geographical structure (such as soil type, climatic conditions). The soil texture, crop type, light or heavy soil, weather are determining factors in irrigation criteria [34]. With tensiometers that measure the real-time soil water potential, crop response can be measured at various levels and applicable water-yield relationships can be developed for optimum crop productivity [35]. In [35], it has been recommended to develop crops that resist the stress factor of water deficit affecting plant growth to increase yield. For the foreseeable future, the variation of water resources according to various parameters should be considered [36, 37]. Efficient, cheap and environmentally acceptable reclamation methods should be developed to improve crop production [38]. Transferring drainage water to suitable areas with correct irrigation management will prevent excessive water usage [39]. The most common strategies adopted for the long-term and efficient use of water resources can be summarized mentioned as in [2].

- Integrated soil and water resources should be planned effectively.
- Irrigation supply systems that will encourage more efficient water use and production should be improved.
- Water allocation policies that support water conservation should be adopted
- Non-beneficial water usage should be reduced.
- Crop restrictions and optimum irrigation methods should be preferred in areas using waste-water.
- Measures should be planned to increase existing resources, including the reuse of waste-water and drainage water.
- Irrigation technologies that contribute to the prevention of water wastes and losses should be adopted.
- User awareness should be created on water resources management.
- Since interruptions in natural water supply occur during droughts, irrigation system management should be changed in addition to water delivery policies.
- Evapotranspiration caused by soil evaporation, seepage and weeds receiving excess irrigation water should be controlled.
- In irrigated agriculture, salinity control must be realized and crops must be managed appropriately to prevent soil degradation.
- The environmental balance should be restored in the use of natural resources for human-induced desertification and water scarcity.
- Water wastes must be minimized and water quality must be managed.

**Table 1:** Irrigation system models.

| Reference | Parameters | Relationships | Application Areas |
|---|---|---|---|
| | radiation, soil heat, air density, | | |
| | air vapour pressure, | crop water requirement and | evapotranspiration based analysis |
| [18] | air heat, psychometric constant, | effective precipitation | |
| | surface and aerodynamic resistances | | |
| [19] | inflow term, temperature, | soil moisture and | wireless sensor network applications |
| | soil moisture, water usage | applied water | |
| | irrigation volume, field capacity, | water, crop transpiration, | |
| [20] | effective rainfall, | soil moisture and | Internet of things applications |
| | groundwater recharge, crop transpiration | lagging irrigation decisions | |
| | economical aspect, | efficiency of seasonal | |
| | knowledge about yield responding to water, | irrigation system and | crop simulation model for |
| [22, 23] | irrigation scheduling, | economical aspect | alternating irrigation conditions |
| | water resources and its limitations | | |
| | pricing, production cost, | crop water allocation, | crop water allocation model |
| [24] | irrigation cost and maximum yields, | economical aspect | design for chosen land area |
| | usage of water | | |
| [25] | hydrodynamic basin | basin irrigation modeling | basin irrigation system applications |
| | irrigation model parameters | | |
| | rainfall, irrigation, surface runoff, | | |
| [21] | crop evapotranspiration, | minimum irrigation water requirement, | Internet of things applications |
| | ground water outflow, infiltration | water balance method | |
| | crop evapotranspiration, | root zone soil | Weather forecast |
| [26] | effective irrigation amount | moisture deficit | applications |
| | effective rainfall | irrigation amount | |
| | climatological, rainfall | | |
| [27] | crop yield, deep percolation | water balance | irrigation scheduling |
| | crop stress factor, | approach | applications |
| | soil moisture depletion | | |
| | evapotranspiration | | |
| | soil moisture, crop yield | soil moisture | seasonal furrow |
| [28] | crop growing, evapotranspiration | hydraulic model | irrigation system |
| | crop stress factor, | crop yield | applications |
| | crop sale, irrigation cost | economic optimization | |
| | soil data, crop data | crop yield | economic |
| [29] | climatic daily data | production | optimization based |
| | economic data | gross margin | water management |
| | water stress | irrigation depth | |
| | hydrological parameters | pond hydrological | pond-irrigation |
| [30] | pond parameters | process and | modeling |
| | evaporation, solar radiation | irrigation demand | |